%&amstex
%Paper: 9202088
%From: gottlieb@gauss.math.purdue.edu (Daniel H. Gottlieb)
%Date: Wed, 26 Feb 92 14:39:07 EST

% amstex 1.1d
%\magnification=\magstep1
\parskip=6pt
\input amstex
\documentstyle{amsppt}
\NoBlackBoxes
\def\TI{\text{Ind}}
\def\CL{\text{cl}}
\document
\null
\vskip 2truein
\centerline{\bf THE INDEX OF DISCONTINUOUS VECTOR FIELDS:}
\centerline{\bf TOPOLOGICAL PARTICLES AND RADIATION}
\medskip
\centerline{by}
\bigskip
\centerline{\bf Daniel H.\ Gottlieb and Geetha Samaranayake}
\bigskip
\centerline{\bf Department of Mathematics}
\bigskip
\centerline{\bf Purdue University}
\centerline{\bf  West Lafayette, Indiana}
\vskip .25truein
\noindent
{\bf Abstract}

We define the concepts of topological particles and topological radiation.
These
are nothing more than connected components of defects of a vector field. To
each
topological particle we assign an index which is an integer which is conserved
under interactions with other particles much as electric charge is conserved.
For space-like vector fields of space-times this index is invariant under all
coordinate transformations. We propose the
following physical principal: {\it For physical vector fields the index
changes only when there is radiation.} As an implication of this principal we
predict that any physical psuedo-vector field has index zero.

\vfill\eject
\baselineskip=18pt
\noindent
{\bf Introduction}

At the frontier between the continuous and the discrete there is a
naturally occurring additive, integral ``quantum number" which is
preserved under ``collisions" of discontinuities. This quantum number
depends only on the basic topological notions of compactness,
connectedness, dimension, and the concept of pointing inside.

We assume we are in a smooth manifold $N$.  A vector field is an
assignment of tangent vectors to some, not necessarily all, of the
points
of $N$.  We make no assumptions about continuity.  We will call this $N$
the {\it arena} for our vector fields.  We consider the set of defects
of a
vector field $V$ in $N$, that is the set $D$ which is the closure of the
set of all zeros, discontinuities and undefined points of $V$.  That is
we
consider a {\it defect} to be a point of $N$ at which $V$ is either not
defined, or is discontinuous, or is the zero vector, or which contains
one
of those points in every neighborhood.

We are interested in the connected components of the defects and how
they
change in time. Those connected components of  $D$  which are compact we
will call topological {\it particles}. If we can find an open set about
a particle
which does not intersect any defect not in the particle itself, then we
say the particle is {\it isolated}. If  $C$  is an isolated particle
we
can assign an integer which we call the index of  $C$  in  $V$.
We denote this by
$\TI(C)$.

The key properties of $\TI(C)$ are that it is nontrivial, additive over
particles, easy to calculate
and is conserved under interactions with proper components as $V$ varies
under
time.  For example, let $V$ be the electric vector field generated by
one
electron in $R^3$.  Then the position of the electron $e$ is the only
defect
and $\TI(e)=-1$.  Now if $V$ changes under time in such a way that there
are only a finite number of particles at each time, all contained
in some large fixed sphere, then the sum of the indices of the particles
at
each time $t$ is equal to $-1$.  Thus the electron vector field can
change to
the proton vector field only if the set of defects changing under time
is unbounded, since the proton has index $+1$ which is different from
the index of the electron. In this case we will say that the
transformation of the electron to the proton involves `` topological
radiation".

Vector fields varying under time, and defect components
interacting with each other, can be made precise by introducing the
concept
of {\it otopy}, which is a generalization of the concept of homotopy.
An
{\it otopy} is a vector field on $N\times I$ so that each vector is
tangent
to a slice $N\times t$.  Thus an otopy is a vector field $W$ on $N\times
I$ so
that $W(n,t)$ is tangent to $N\times t$.  We say that $V_0$ is otopic to
$V_1$
if
$V_0(n)=W(n,0)$ and $V_1(n)=W(n,1)$. We say that a set of components
$C_i$
of defects on $V_0$ {\it transforms} into a set of components of
defects $D_j$
of
$V_1$ if there is a connected component $T$ of the defects of $W$ so
that
$T\cap (N\times 0)=\cup C_i$ and $T\cap(N\times 1)=\cup D_j$.  If $T$ is
a
compact connected component of defects of $W$, which transforms a set of
isolated particles $C_i$ into isolated particles  $D_j$,  then we say
there is no topological {\it radiation} and
$$
\sum \TI(C_i)=\sum \TI(D_j).\tag1
$$
If $T$ is not compact, we say there is topological radiation.

We define $\TI(C)$ as follows.  Since $C$ is an particle, there is an
open set
$U$ containing $C$ so that there are no
defects in the closure of $U$ except for $C$.  We can define an index
for
any vector field defined on the closure of an open set so that the set
of
defects is compact and there is no defect on the frontier of the open
set.
We say such a vector field is {\it proper} with {\it domain} the open
set.
In the case at hand, $V$ restricted to $\CL(U)$ is proper with domain
$U$.
Hence we can define $\TI(V|U)$.  We set $\TI(C)=\TI(V|U)$.

Next we define $\TI(V)$ with domain $U$ to be equal to the index of
$V|M$ where
$M\subset U$ is a smooth compact manifold with boundary containing the
defects
of $V$ in its interior.  We can find such an $M$ since the defects are a
compact set in $U$.

We call a vector field $V$ defined on a compact manifold $M$ {\it
proper}
if there are no defects on the boundary.  Consider the open set of the
boundary where $V$ points inside.  We denote that set by $\partial_-M$.
We define the vector field $\partial_-V$ with domain $\partial_-M$ in
the
arena $\partial M$ by letting $\partial_-V$ be the end product of first
restricting $V$ to the boundary and then projecting each vector so that
it
is tangent to $\partial M$ which results in a vector field $\partial V$
tangent
to $\partial M$, and then finally restricting $\partial V$ to
$\partial_-M$
to get $\partial_-V$. Then we define $\TI(V)$ by the equation
$$
\TI(V)=\chi(M)-\TI(\partial_-V)\tag*
$$
where $\chi(M)$ denotes the Euler-Poincare number of $M$.  We know that
$\partial_-V$ is a proper vector field with domain $\partial_-M$ since
the
set of defects is compact unless there is a defect at the the frontier
of
$\partial_-M$.  If there were such a defect, it would be a zero of $V$
tangent
to $\partial M$ and hence a zero of $V$ on the boundary, so $V$ would
not have
been proper.

Now $\partial_-V$ is a proper vector field with domain the open set
$\partial_-M$ which is one dimension lower than $M$. Then $\TI (
\partial_-V)$
is defined in turn by finding a compact manifold containing the defects
of
$\partial_-V$ and using equation (*). We continue this process until
either $\partial_-M$ is a zero dimensional manifold where every point is
a
defect and so $\TI(\partial_-V)$ is simply the number of points, or where
$\partial_-M$ empty in which case $\TI(\partial_-V)=0$.

To summarize, we define the index of a proper vector field $V$ with
domain $U$
assuming
that the index for vector fields is already defined for compact
manifolds
with boundary.  Then the index of $V$ is defined to be the index of $V$
restricted to a compact smooth manifold with boundary of codimension
zero
containing
all the defects of $V$ in $U$.  We will show this definition is
well-defined,
that
is it does not depend on the chosen manifold with boundary, by showing
that a vector field with no defects defined on a compact manifold with
boundary has index zero.

The well-definedness of this definition will involve the first four
sections
of this paper. In section 5 we summarize the useful properties of the
index
which we have proved along the way, along with a few proved in other
papers.
The key property is that of a {\it proper otopy} described below.

Suppose that $V$ is a proper vector field with open domain $U$.  A {\it
proper
otopy} is a proper vector field $W$
defined on $N\times I$ with domain an open set where we require $W$ to
be
tangent to the slices.  Then we say $W$ is a proper otopy of $V$ if $V$
is
the restriction of $W$ to $N\times 0$ and the domain of $W$ intersects
$N\times 0$ in $U$.  The key property of the index of proper vector
fields
with open domains is that the index is invariant under proper otopy. For
connected manifolds the converse is true: Two proper vector fields are
properly otopic if and only if they have the same index.

We may generalize the concept of otopy in two ways.  Recall an otopy is
an
open set $T$ on $N\times I$ with a vector field $W$ which is tangent to
the
slices.  Now this can be generalized by considering a fibre bundle
$E\to B$ with fibre $N$ and an open set $T$ on $E$ and a vector field
$W$ whose vectors are tangent to the fibre.  It is clear that if $W$ is
a
proper vector field, that is the defects form a compact set and there
are no
defects on the frontier of $T$, then $W$ restricted to any fibre has an
index.
This index is the same for every fibre.  In [B-G], for the case of
continuous
$W$, it is shown that there is an $S$-map which induces a transfer on
homology
with trace equal to this index.

The second way to generalize an otopy is to note that $N\times I$ can be
thought of as a manifold $S$ with a natural non-zero vector field.  Then
$W$ is a vector field which is orthogonal to this vector field.  In fact
any
vector field can be projected orthogonal to the  natural vector field.
If
$S$ is a space-time, there is a field of light cones.  If we consider
a space-like vector field $W$ on $S$, it is like an otopy.  $W$
restricts
to any space-like slice and projects tangent to it.  The index of the
defects
at any event is thus an invariant of general relativity, it is invariant
under any change of coordinate system.  The defects propagate through
space-time and the index satisfies a conservation law, just like the
conservation law of electric charges under particle collisions.  It is
very easy to believe
that the index of a vector field, as here exposed, must lead to an
explanation of the conservation of physical properties under collision
based
on the idea of connectivity and continuity and pointing inside.

As a first step in this direction we make the following proposal. Every
physical vector field for which the index is defined,
has the same index under any
choice of coordinates and orientation. Hence we conjecture that any
psuedo vector field must have either the index equal to zero or
the index undefined. Also we propose that whenever a physical vector
field has a change in its index, then there must have been radiation.
\bigskip

\noindent
{\bf 1.  The definition for one-dimensional manifolds}
\medskip

The inductive definition begins with empty vector fields, that is domains
which are empty.  This could arise since $\partial_-M$ is empty if $V$
never points inside from the boundary.  We define the index of an empty
vector field to be equal to zero.  Zero dimensional manifolds consist of
discrete sets of points.  The only vectors are zero vectors, so for a
vector field to be proper it must consist of a finite number of zeros.
One-dimensional compact manifolds with boundary consist of a finite disjoint
union of compact components which are compact intervals.  We use the
definition (*), that is
$$
\aligned
\TI(V)&=\text{ (number of components) \text{--} (number of boundary}\\
&\qquad\text{points where $V$ is pointing inwards)}.
\endaligned
$$
In the case of components without boundaries, circles in this case, we
define the index to be $\chi(\text{circle})=0$.

\proclaim{Lemma 1.1}
Two vector fields $V$ and $V'$ are properly otopic if and only if
$$
\TI(\partial_-V)=\TI(\partial_-V')\text{ on each component of the boundary}.
$$
\endproclaim

\demo{Proof}
Let $W$ be a vector field so that $W(m)=V(m)/\Vert V(m)\Vert$ for $m$ on
the boundary of $M$.  Assume that $W(m)=0$ outside a collar of the
boundary, and assume that $W$ continuously decreases in size from the unit
vectors on the boundary to the zero vectors at the other end of the collar.
Then we define the homotopy $tV+(1-t)W$.  This is a proper homotopy, since
at any point $m$ on the boundary  $V(m)$ and $W(m)$ both point either inside
or outside so no zero can arise on the boundary.  If $V$ should have a defect
at some $m$ in the interior, we may alter $V$ by assigning $V(m)=0$.  Thus the
homotopy is defined.  Now both $V$ and $V'$ are properly otopic to $W$, hence
they are otopic to each other.
\enddemo

\proclaim{Lemma 1.2}
If $M$ is a finite collection of manifolds with boundary and $f$ is a
diffeomorphism so that the related vector field is denoted by $V^*$, then
$$
\TI(V)=\TI(V^*).
$$
\endproclaim

\demo{Proof}
Pointing inside is preserved under diffeomorphism.
\enddemo

\proclaim{Lemma 1.3}
If $V$ has no defects, then $\TI(V)=0$.
\endproclaim

\demo{Proof}
Each connected component of $M$ is an interval.  Since $V$ has no defects
on this interval, $V$ must point outside on one end and inside on the
other.  Thus $\TI(V)=1-1=0$ on this interval, and thus on all the intervals.
So $\TI(V)=0$ is true for $M$.

Now suppose that the arena is a connected manifold $N$ with no boundary
and not compact.   Thus an open interval.  Then we define the index of $V$
with open domains to be the index of $V$ restricted to a union of compact
intervals which contain the defects of $V$.  This is well-defined.  If $M$
and $M'$ are two manifolds with boundary containing the defects, there is a
compact manifold with boundary $M''$ containing both $M$ and $M'$.  The
vector field $V$ restricted to $M''-\text{int}(M)$ is a nowhere zero vector
field, and the previous lemma and the fact that the index is additive proves
that the index is well-defined.

Next we deal with the case of the arena $N$ being a closed manifold, in this
case that is a finite set of circles.  We will consider the case of a single
circle, the general case will be given by adding the indices for each
connected component.  The set of defects is closed.  If the defects can be
contained in a compact manifold with boundary, in this case diffeomorphic
to a closed interval, we define the index of $V$ to be the index of $V$
restricted to the compact manifold.  On the other hand, if the domain of $V$ is
the
entire arena, then we define
$$
\TI(V)=\chi(\text{arena})-\TI(\partial_-V)=\chi(\text{circle})-
\TI(\text{empty vector field})=0.
$$

These two definitions are consistent. If $V$ has domain the entire circle,
then it is properly homotopic to the zero vector field.  Then we homotopic
the zero vector field to $V'$ which is zero inside a large closed interval
and not zero around a point with the vectors thus forced to point in the same
sense around the circle.  Then $V'$ restricted to the large closed interval
has index zero which is just what the global definition gives.

We make a few more observations before we finish with the one-dimensional
case.
\enddemo

\proclaim{Lemma 1.4}
Given a connected arena $N$, two proper vector fields are properly otopic
if and only if they have the same index.  For every integer $n$ there is a
vector field whose index equals that integer.
\endproclaim

\demo{Proof}
Suppose we have a proper otopy $W$ with domain $T$ on $N\times I$.  Let $V_t$
denote $W$ restricted to $N\times t$.  We show that there is some interval
about $t$ such that $V_s$ has the same index for all $s$ in the interval.
Since the set of defects of the otopy is compact we can find a compact manifold
$M$ so that $M\times J$, for some closed interval $J$, lies in $T$ and contains
the defects inside $\partial M\times J$.  Thus the proper homotopy $V_t$ on
$M\times J$ preserves the index on $M$, and hence the proper otopy on
$N\times J$ preserves the index on $N$ as $t$ runs over $J$.  Thus we have a
finite sequence of vector fields each having the same index as the previous
vector field.  Hence the first and last vector fields have equal indices.
Conversely, for any integer $n$, let $W_n$ be the vector field consisting
of $|n|$ vector fields defined on disjoint open intervals in $N$, each one of
index $1$ if $n>0$ and of index $-1$ if $n<0$.  Thus $\TI(W_n)=n$.  Now if
$V$ has index $n$, we must show that $V$ is properly homotopic to $W_n$.  Now
the domain of $V$ consists of open connected intervals, and only a finite
number of them contain defects.  Each of these intervals has index equal to
$1$, $-1$, or $0$.  Now $V$ is properly otopic to the same vector field $V$
whose domain is restricted to only those intervals which have nonzero indices.
Now if two adjacent intervals have different indices, there is a proper otopy
which leaves the rest of the vector field fixed, and removes the two
intervals of opposite indices.  After a finite number of steps we are left
with either an empty vector field, if $n=0$, or a $W_n$.  The empty vector
field is $W_0$.  Thus $V$ is properly otopic to $W_n$.
\enddemo

\proclaim{Lemma 1.5}
The index of a vector field on an open manifold is invariant under
diffeomorphism.
\endproclaim

\demo{Proof}
Immediate from Lemma 1.2 and the definition of index for open manifolds.
\enddemo

\proclaim{Lemma 1.6}
Let $V$ be a vector field over a domain $U$ and suppose that $U$ is the
disjoint union of $U_1$ and $U_2$.  Then if $V_1$ and $V_2$ denote $V$
restricted to $U_1$ and $U_2$ respectively, we have
$$
\TI(V)=\TI(V_1)+\TI(V_2).
$$
\endproclaim
\medskip
\noindent
{\bf 2.  The index defined for compact $n$-manifolds}

\proclaim {The otopy extension property}
Let $V$ be a continuous vector field on a closed manifold $N$.  Let $U$ be an
open set in $N$.  Any continuous
proper otopy of $V$ on the domain $U$ can be extended to a continuous
homotopy of $V$ on all of $N$.
\endproclaim
\demo{Proof}
The continuous proper otopy implies there is a continuous vector field $W$
on an open set $T$ in $N\times I$ which extends to the closure of $T$ with
no zeros on the frontier and which is $V$ when restricted to $N\times 0$.
This vector field $W$ can be thought of as a cross-section to the tangent
bundle over $N\times I$ defined over a closed subset. It is well known that
cross-sections can be extended from closed sets to continuous cross-sections
over the whole manifold.
\enddemo

We assume that the index is defined for $(n-1)$-manifolds in such a way that
all the lemmas of section 1 hold.

First we consider the case of compact manifolds such that every component is
a manifold with boundary.  We suppose that $V$ is a proper vector field on
such a manifold $M$.  We choose a vector field $N$ on the boundary
$\partial M$ which points outside of $M$.  Every vector $v$ at a point $m$
on $\partial M$ can be uniquely written as $v=t+kN(m)$ where $t$ is a vector
tangent to $\partial M$ and $k$ is some real number.  We say $t$ is the
{\it projection} of $v$ tangent to $\partial M$.  Then $\partial V$ is the
vector field obtained by projecting $V$ tangent to $\partial M$.  Now we
define $\partial_-V$ by restricting $\partial V$ to $\partial_-M$, the set
of points such that $V$ is pointing inward.  Then we define
$$
\TI(V)=\chi(M)-\TI(\partial_-V).\tag*
$$

\proclaim{Lemma 2.1}
$\TI(V)$ is well-defined.
\endproclaim

\demo{Proof}
We have already defined the index on $(n-1)$-dimensional manifolds with
open domains for proper vector fields.  Note that $\partial_-V$ is proper
since $V$ is, since the frontier of $\partial_-M$ is a subset of
$\partial_0M$ where $V$ is tangent to $\partial M$.  So a defect of
$\partial_-V$ on the frontier must come from a defect of $V$ on $\partial M$.
Hence $\TI(\partial_-V)$ is defined.  Now the vector field $\partial_-V$
obviously depends upon the outward pointing $N$.  If we had another outward
pointing vector field $N'$ we would project down to a different
$\partial_-V$, call it $W$.  Now the homotopy of vector fields
$N_t=tN+(t-1)N'$ always points outside of $M$ for every $t$.  Hence it
induces a homotopy from $\partial_-V$ to $W$ and this homotopy is proper.
Thus $\TI(\partial_-V)=\TI(W)$.
\enddemo

We will also allow the case where $N$ is not defined on a closed set of
$\partial M$ which is disjoint from the frontier of
$\partial_-M$.  Then $\partial V$ has
defects, but $\partial_-V$ is still proper.  A homotopy between $N$ and
$N'$, as in the lemma, still induces a proper otopy between $\partial_-V$
and $W$, so the $\TI(V)$ is still well-defined in this case also.  This
case arises when $M$ is embedded as a co-dimension zero manifold in such a
way that it has corners.  Then the natural outward pointing normal in this
situation is not defined on the corners.  But we still have the index defined
if none of the corners is on the frontier of $\partial_-M$.

Now our goal is to prove that non-zero vector fields have index equal to
zero on compact manifolds with boundary.

\proclaim{Theorem 2.2}
$V$ is properly otopic to $W$ if and only if
$$
\TI(\partial_-V)=\TI(\partial_-W)
$$
for every connected component of $\partial M$.  So as a corollary in the
case that $\partial M$ is connected, we have that $V$ is properly otopic
to $W$ if and only if $\TI(V)=\TI(W)$.  If $V$ and $W$ are both continuous,
then ``otopic'' can be replaced by ``homotopic'' in the above statements.
\endproclaim

\demo{Proof}
The theorem is true for manifolds one dimension lower by lemma 1.1.  A proper
otopy of $V$ to $W$ induces a proper otopy from $\partial_-V$ to
$\partial_-W$ in the arena $\partial M$.  Hence $\TI(\partial_-V)=
\TI(\partial_-W)$.  Hence $\TI(V)=\TI(W)$ from $(*)$.  Conversely, we can
find a smooth collar $\partial M\times I$ of the boundary so that $V$
restricted to this collar has no defects.  Then we otopy $V$ to $V'$ where
$V'$ is defined by $V'(m,t)=tV(m)$ for a point in the collar and $V'=0$ outside
the collar.  Now since $\TI(\partial_-V)=\TI(\partial_-W)$ for each
connected component of the boundary, we can find a proper otopy from
$\partial_-V$ to $\partial_-W$.  Now this otopy can be extended to a homotopy
of $\partial V$ to $\partial W$ by the otopy extension property.  This
homotopy in turn can be used to define a proper homotopy from $V'$ to $W'$.
Here we assume $W'$ has the same definition relative to $W$ as $V'$ has to $V$.
Thus $W$ is properly otopic to $V$.
\enddemo

\proclaim{Lemma 2.3}
Suppose $V$ is a proper vector field on a compact manifold $M$ each of whose
components has a non-empty boundary.  Let $\partial M\times I$ be a collar
of the boundary so small so that $V$ has no defects on the collar.  Then $V$
restricted to $M$ minus the open collar $\partial M\times (0,1]$ has the same
index as $V$.
\endproclaim

\demo{Proof}
Let $\partial V_t$ denote the projection of $V$ tangent to the submanifold
$\partial M\times t$ for every $t$ in $I$.  Let $W$ be the vector field on
the collar defined by $W(m,t)=\partial_-V_t$ if $(m,t)$ is a point in
$\partial_-M\times t$.  Then $W$ is a proper otopy, proper since $V$ has no
defects on the collar.  Thus $\TI(\partial_-V)=\TI(\partial_-V_0)$ and
hence $\TI(V)=\chi(M)-\TI(\partial_-V)$ equals the index of $V$ restricted
to $M'=M- open\ collar$, because the indices of the $\partial_-$
vector fields are the same on their respective boundaries and $\chi(M)=
\chi(M')$.
\enddemo

\proclaim{Lemma 2.4}
Let $V$ be a proper continuous vector field on $M$.  Suppose that $\partial_-V$
is properly otopic to some vector field $W$ on $\partial M$.  Then there
is a proper homotopy of $V$ to a proper continuous vector field $X$ so that
$\partial_-X=W$ and the zeros of each stage of the homotopy $V_t$ are not
changed.
\endproclaim

\demo{Proof}
Use the otopy extension property to find a homotopy $H_t$ from $\partial V$
to a vector field on $\partial M$ which we shall call $\partial X$. Let
$n(m,t)$ be a continuous real valued function on $\partial M\times I$ which
is positive on the open set $T$ of the otopy between $\partial_-V$ and $W$,
zero on the frontier of $T$, and negative in the complement of the closure
of $T$, and so that $n(m,0)=n(m)$ where $V(m)=n(m)N(m)+\partial V(m)$
defines $n(m)$.  Such a function exists by the Tietze extension theorem.
Using $n(m,t)$, we define a vector field $X'$ on $\partial M\times I$ by
$X'(m,t)=n(m,t)N(m)+H_t(m)$.  We adjoin the collar to $M$ as an external
collar and extend the vector field $V$ by $X'$ to get the continuous vector
field $X$.  Now $M$ with the external collar is diffeomorphic to $M$.  Under
this diffeomorphism $X$ becomes a vector field which we still denote by $X$.
We may assume this diffeomorphism was so chosen that $X=V$ outside of a small
internal collar.  Then the homotopy $tX+(1-t)V$ is the required homotopy which
does not change the zeros of $V$.
\enddemo

\proclaim{Lemma 2.5}
If $V$ is a vector field with no defects on an $n$-ball, then $\TI(V)=0$.
\endproclaim

\demo{Proof}
For the standard $n$-ball of radius $1$ and center at the origin, we define
the homotopy $W_t(r)=V(tr)$. This homotopy introduces no zeros and shows
that $V$ is homotopic to the constant vector field.  The constant vector
field has index equal to zero, as can be seen by using  $(*)$.  If we have
a ball diffeomorphic to the standard ball, then the index of the vector field
under the diffeomorphism is preserved, and hence it has the zero index.  If
the ball is embedded with corners so that the corners are not on the
frontier of the set of inward pointing vectors of $V$, then the index is
defined
and by lemma 2.3 it is equal to the index of $V$ restricted to a smooth ball
slightly inside the original ball.  This index is zero.
\enddemo

\proclaim{Theorem 2.6}
If $V$ is a vector field with no defects on a compact manifold such that all
the components have non-empty boundary, then $\TI(V)=0$.
\endproclaim

\demo{Proof}
Now $M$ can be triangulated and suppose we have proved the theorem for
manifolds triangulated by $k-1$ $n$-simplicies.  The previous lemma proves
the case $k=1$.  We divide $M$ by a manifold $L$ of one lower dimension
into manifolds $M_1$ and $M_2$ each covered by fewer than $k$ $n$-simplicies
so that the theorem holds for them.

We arrange it so that $L$ is orthogonal to $\partial M$.  We use lemma 2.4 to
homotopy $V$ to a vector field with no defects so that the new $V$ is pointing
outside orthogonally to $\partial M$ at $L\cap\partial M$. Then a simple
counting argument shows that $\TI(V)=0$ since the restrictions of $V$ to $M_1$
and $M_2$ have index zero.  This argument works if $M$ has no corners.  If
$M$ has corners we find a collar of $M$ which is a smooth embedding of
$\partial M\times t$ for all $t$ but the last $t=1$. Then by lemma 2.3 above,
we
find that $V$, restricted to the manifold bounded by $\partial M\times t$ for
$t$ close enough to $1$, has the same index as $V$.  That is zero.

The counting argument goes as follows.  By induction,
$\TI(V|M_1)=\TI(V|M_2)=0$.
Thus $\TI(\partial_-V_1)=\chi(M_1)$ and $\TI(\partial_-V_2)=\chi(M_2)$.
Now $\TI(\partial_-V)=\TI(\partial_-V_1)+\TI(\partial_-V_2)-\TI(W)$ where
$W$ is the projection of $V$ on the common part of the boundary of $M_1$ and
$M_2$, that is $L$.  This follows from repeated applications of lemma 1.6.
Now $\TI(W)=\chi(L)$ since $W$ points outwards at the boundary of $L$.  Hence
$$
\TI(\partial_-V)=\TI(\partial_-V_1)+\TI(\partial_-V_2)-\TI(W)=
\chi(M_1)+\chi(M_2)-\chi(L)=\chi(M).
$$
Hence $\TI(V)=0$ from $(*)$.
\enddemo

\medskip
\noindent
{\bf 3.  The index for open $n$-manifolds}

Let $N$ be an $n$-manifold and let $V$ be a proper vector field on $N$ with
domain $U$.  Then the set of defects of $V$ in $U$ is compact.  Thus we can
find a compact manifold $M$ which contains the defects of $V$.  We define
$$
\TI(V)=\TI(V|M).\tag**
$$

\proclaim{Lemma 3.1}
$\TI(V)$ is well-defined.
\endproclaim

\demo{Proof}
If $M$ and $M'$ are two manifolds with boundary containing the defects, there
is a compact manifold with boundary $M''$ containing both $M$ and $M'$.  The
vector field $V$ restricted to $M''-\text{int}(M)$ is a nowhere zero vector
field.  Then Theorem 2.6 implies that the index of $V$ restricted to
$M''-\text{int}(M)$ is zero.  Now the index of $V$ restricted to $M''$ equals
the index of $V$ restricted to $M$ by the following lemma.
\enddemo

\proclaim{Lemma 3.2}
Suppose $M$ is the union of two manifolds $M_1$ and $M_2$ where the three
manifolds are compact manifolds with boundary so that the intersection of
$M_1$ and $M_2$ consist of part of the boundary of $M_1$ and is disjoint
{}from the boundary of $M$.  Suppose that $V$ is a proper vector field defined
on $M$ which has no defects on the boundaries of $M_1$ and $M_2$.  Then
$\TI(V)=\TI(V_1)+\TI(V_2)$ where $V_i=V|M_i$.
\endproclaim

\demo{Proof}
$$
\aligned
\TI(V) &= \chi(M)-\TI(\partial_-V)\\
&= \chi(M)-(\TI(\partial_-V_1)+\TI(\partial_-V_2)-\TI(\partial_-V_1|L)-\TI(
\partial_-V_2|L))
\endaligned
$$
where $L=M_1\cap M_2$.  Now
$$
%% FOLLOWING LINE CANNOT BE BROKEN BEFORE 80 CHAR
\TI(\partial_-V_1|L)+\TI(\partial_-V_2|L)=\TI(\partial_-V_1|L)+\TI(\partial_+V_1)=\chi(L).
$$
Thus
$$
\TI(V)=\chi(M_1)+\chi(M_2)-\TI(\partial_-V_1)-\TI(\partial_-V_2)=
\TI(V_1)+\TI(V_2),
$$
as was to be proved.
\enddemo

\proclaim{Lemma 3.3}
Let $V$ be a proper vector field with domain $U$.  Suppose $U$ is the union
of two open sets $U_1$ and $U_2$ such that the restriction of $V$ to each of
them and to $U_1\cap U_2$ is a proper vector field denoted $V_1$ and
$V_2$ and $V_{12}$ respectively. Then
$$
\TI(V)=\TI(V_1)+\TI(V_2)-\TI(V_{12}).\tag***
$$
\endproclaim

\demo{Proof}
We choose disjoint compact manifolds $M_1$, $M_2$, and $M_{12}$ containing
the zeros of $V$ which lie in $U_1-U_{12}$ and $U_2-U_{12}$ and $U_{12}$
respectively.  Then the index of $V$ is equal to the index of $V$ restricted
to the union of $M_1$, $M_2$, and $M_{12}$.  But the index of $V_1$ is the
index of $V$ restricted to $M_1$ and $M_{12}$, and the index of $V_2$ is the
index of $V$ restricted to $M_2$ and $M_{12}$, and the index of $V_{12}$ is
the index of $V$ restricted to $M_{12}$.  Hence counting the index gives the
equation $(***)$.
\enddemo

\proclaim{Theorem 3.4}
Given a connected arena $N$, two proper vector fields are propely otopic if
and only if they have the same index.  For every integer $n$ there is a vector
field whose index equals that integer.
\endproclaim

\demo{Proof}
Suppose we have a proper otopy $W$ with domain $T$ on $N\times I$.  Let
$V_t$ denote $W$ restricted to $N\times t$.  We show that there is some
interval about $t$ such that $V_s$ has the same index for all $s$ in the
interval.  Since the set of defects of the otopy is compact we can find a
compact manifold $M$ so that $M\times J$, for some closed interval $J$, lies
in $T$ and contains the defects so that the defects avoid
$\partial M\times J$.  Thus the
proper homotopy $V_t$ on $M\times J$ preserves the index on $M$, and hence
the proper otopy on $N\times J$ preserves the index on $N$ as $t$ runs over
$J$.  Thus we have a finite sequence of vector fields each having the same
index as the previous vector field.  Hence the first and last vector fields
have equal indices.

Conversely, for any integer $k$, let $W_k$ be the vector field consisting of
$|k|$ vector fields defined on disjoint open balls in $N$, each one of index
$1$ if $k>0$ or of index $-1$ if $k<0$.  Thus $\TI(W_k)=k$.  Now if $V$ has
index $k$, we must show that $V$ is properly homotopic to $W_k$.  Now the
defects of $V$ form a compact set which are contained in a compact manifold
with boundary $M$ so that $V$ is defined and has no defects on the boundary.
We may proper otopy $V$ first to a continuous vector field, and then to a
smooth
vector field.  Then we consider $V$ as a cross-section to the tangent
bundle of $M$.  Using the transversality theorem, we can smoothly homotopy
the cross-section so that it is transversal to the zero section of the
tangent bundle keeping the cross-section fixed over the boundary.  The
dimensions are such that the intersection consists of a finite number of
points.  Thus we proper otopy $V$ to a vector field with only a finite number
of zeros.  Now we put small open balls around each of these zeros.  The
index of the vector field on the ball around each of these zeros is either
$1$ or $-1$.  This follows from transversality, but we do not need that
fact.  We may find a diffeomorphic $n$-ball which contains exactly $|k|$
zeros so that around these zeros the vector field restricts to $W_k$.  The
two vector fields have the same index on the $n$-ball and thus are properly
homotopic, since from $(*)$ the index on the boundary of the inward pointing
$\partial_-$ vector fields is the same, and so by induction they are properly
otopic, hence by the otopy extension property the $\partial$ vector fields
are homotopic.  This homotopy can be extended to a homotopy of the two vector
fields originally on the $n$-ball.  Then using the sequence of homotopies
and
otopies, we can piece together a proper otopy of $V$ to $W_k$.
\enddemo

\proclaim{Corollary 3.5}
The proper homotopy classes of continuous proper vector fields on a compact
manifold with connected boundary is in one-to-one correspondence with
the integers via the index.
\endproclaim

\proclaim{Lemma 3.8}
The index of a vector field on an open manifold is invariant under
diffeomorphism.
\endproclaim

\proclaim{Lemma 3.9}
The index of a vector field $V$ on a closed manifold $M$ whose domain is the
whole of $M$ is equal to $\chi(M)$.
\endproclaim

\demo{Proof}
First otopy $V$ to the zero vector field.  Then homotopy the zero vector
field to a vector field $V'$ so that it is a non-zero vector field on a
small $n$-ball $B$ about a point.  Now let $V_1$ be $V'$ on the $n$-ball and
let $V_2$ be $V'$ on the complement.  Then $\TI(V_1)=0$, so
$\TI(\partial_-V_1)=1$.  Now $\TI(\partial_-V_2)=(-1)^{n-1}$.  So
$$
\TI(V_2)=\chi(M-B)-(-1)^{n-1}=\chi(M)-(-1)^n-(-1)^{n-1}=\chi(M).
$$
Hence $\TI(V)=\TI(V_1)+\TI(V_2)=0+\chi(M)$.
\enddemo

\noindent
{\bf 4.  The Index of particles}

Let $V$ be a vector field on an arena $N$.  Let $D$ be the set of defects
of $V$. Then $D$ breaks up into a set of connected components $D_i$. We
define an index for each component $D_i$ which is compact and is an open
set in the subspace topology of $D$. That is, in the terminology of the
Introduction, we define the index of an isolated particle.
For isolated particles we can find a compact
manifold $M$ containing $D_i$ and no other defects.  Then we define
$$
\TI(D_i)=\TI(V|M).\tag****
$$

Now if we have a finite number of particles $D_i$ in the domain of $V$, then
$\TI(V)=\sum_i \TI(D_i)$.  However it is possible that $V$ is a proper
vector field and there are an infinite number of $D_i$.  Then at least one
of the $D_i$ is not isolated in $D$.  But the index of $V$ is still defined.
This event is very rare in practical situations.  A one dimensional example
occurs when $M$ is the interval $[-1,1]$ and the vector field $V$ is defined
by $V(x)=x\sin(1/x)$ for $x\not=0$ and $V(0)=0$.  Then $0$ is a connected
component of the defects which is not open in the set of zeros of $V$.

If we have an otopy $V_t$, we imagine the components of the defects $D_t$ as
changing under time.  We can say that $D_{ti}$ at time $t$ transforms without
radiation into
$D_{sj}$ at time $s$ if there is a compact connected component $T$ of the
defects of the otopy from time $t$ to time $s$ so that $T$ intersects
$N\times t$ in exactly $D_{ti}$ and $T$ intersects
$N\times s$ exactly at $D_{si}$.
The index of $D_{ti}$ is the same as the index of $D_{sj}$ if $T$ is compact.
In other words if a finite number of particles $D_{i}$ at time $t$ are
transformed into a finite
number of particles $C_{j}$ at time $s$ by a compact $T$,
the sum of the indices are
conserved. That is
$$
\sum \TI(C_i)=\sum \TI(D_j).\tag1
$$

Thus the idea of otopy allows us to make precise the concept of defects
moving with time and changing with time and undergoing collisions.  The
index is conserved under these collisions as long as the ``world line'' $T$
of the component is compact. That is, as long as there are is no
radiation.

\bigskip
\noindent
{\bf 5.  Properties of the Index}
\medskip
\noindent
(2)\quad$\TI(V)+\TI\ \partial_-V=\chi(M)$

This is in fact the equation (*) which defines the index.
\medskip
\noindent
(3)\quad Let $N$ be a connected arena.  $V$ is a properly otopic to $W$ if and
only if $\TI\ V=\TI\ W$.  For any integer $n$ there is a vector field $W$ so
that $n=\TI\ W$.
\medskip
\noindent
(4)\quad Suppose $M$ is a compact manifold so that $\partial M$ is connected,
and
suppose $V$ and $W$ are continuous proper vector fields on $M$.  Then
$V$ is properly homotopic to $W$ if and only if $\TI\ V=\TI\ W$.  For any
integer $n$ there is a continuous proper vector field $W$ so that
$n=\TI\ W$.
\medskip

\noindent
(5)\quad If $M$ is a closed compact manifold and $V$ is a vector field whose
domain is all of $M$, then $\TI\ V=\chi(M)$.
\medskip

\demo{Proof}
Property (3) and (4) are Theorem 3.4 and Corollary 3.5 respectively for the
homotopy part.  For the fact that $n=\TI\ W$ for some vector field $W$, we
apply (2) and induction starting with Lemma 1.4.  The proof of (5) is
Lemma 3.9.
\enddemo

\noindent
(6)\quad Let $A$ and $B$ be open sets and let $V$ be a proper vector field on
$A\cup B$ so that $V|A$ and $V|B$ are also proper.  Then $\TI(V|A\cup B)=
\TI(V|A)+\TI(V|B)-\TI(V|A\cap B)$.
\medskip

\demo{Proof of (6)}
Lemma 3.3
\enddemo
\medskip
\noindent
(7)\quad Suppose $V$ us a vector field with no defects.  Then $\TI\ V=0$.
\medskip

\demo{Proof}
Theorem 2.6 for compact manifolds with boundary.
\enddemo

\noindent
(8)\quad Suppose $V$ is a proper vector field and the set of defects consists
of a finite number of connected components $D_i$.  Then
$\TI\ V=\underset i\to\sum\ \TI(D_i)$.
\medskip
\demo{Proof}
This follows from the definition of $\TI(D_i)$ and (3).
\enddemo

\noindent
(9)\quad  Let $V$ and $W$ be proper vector fields on $A$ and $B$ respectively.
Let $V\times W$ be a vector field on $A\times B$ defined by $V\times W(s,t)=
(V(s),W(t))$.  Then $\TI(V\times W)=(\TI\ V)\cdot(\TI\ W)$.
\medskip

\demo{Proof}
We can assume that $A$ and $B$ are open sets in their arenas.  Then $V$ is
otopic to $V_n$ where $V_n$ is restricted to a finite set of open sets
in $A$ homeomorphic to the interior of $I^k$ when $k=\text{ dim }A$ and so
that $V_n(t_1,\dots,t_k)=(\pm t_1,t_2,\dots,t_k)$ where the $+t_1$ is taken
if $\TI\ V$ is positive and $-t_1$ is taken if $\TI\ V$ is negative.  The
index of the $V_n|I_k$ is $\pm 1$ respectively (by induction on (9)).  So
$\TI\ (V\times W)=(\TI\ V_n\times W_n)=\underset {i,j}\to\sum\ \TI(V_n|I_i^k)
\times(W_n|I_j^\ell)$.  Now it is easy to see that $\TI(V_n|I^k_i)\times
(W_n|I_j^\ell))=\TI(V_n|I_i^k)\cdot\TI(W_n|I_j^k))$.
\enddemo

\noindent
(10)\quad $(-1)^n\TI(V)=\TI(-V)$ where $n=\text{ dim } M$.
\medskip

\demo{Proof}
The theorem is true for $n=1$.  Assume it is true for $(n-1)$-manifolds.
Now using (2) we have
$$
\aligned
\TI(-V) &=\chi(M)-\TI(\partial_-(-V))\quad \text{ by (2) }\\
&=\chi(M) -\TI(-\partial_+V)\quad \text{by definition of $\partial_-V$ and
$\partial_+V$}\\
&=\chi(M)-(-1)^{n-1}\TI(\partial_+(V))\quad \text{by induction}\\
&=\chi(M)+(-1)^n(\chi(\partial M)-\TI(\partial_-V))
\endaligned
$$
since
$$
\chi(\partial M)=\TI(\partial_-V)+\TI(\partial_+V).
$$
If $n$ is even then
$$
\TI(-V)=\chi(M)+(0-\TI(\partial_-V))=\TI\ V\quad \text{ by (2)}.
$$
If $n$ is odd then
$$
\aligned
\TI(-V) &=\chi(M)-(2\chi(M)-\TI(\partial_-V))\\
&=-(\chi(M)-\TI(\partial_-V))=-\TI\ V\quad \text{ by (2)}
\endaligned
$$
\enddemo

\noindent
(11)\quad Suppose $M$ is a compact sub-manifold of ${\Bbb R}^n$ of
$0$-codimension.  Let $f:M\to {\Bbb R}^n$ be a map so that $f(\partial M)$ does
not
contain the origin.  Define a proper vector field $V^f$ on $M$ by
$V^f(m)=f(m)$.
Then $\TI\ V^f=deg\ f'$ where $f':\partial M\to S^{n-1}$ by
$f'(m)=\frac{f(m)}{\Vert f(m)\Vert}$.

\demo{Proof}
We homotopy $f$ if necessary so that $\vec 0$ is a regular value.  Then
$f^{-1}(\vec 0)$ is a finite set of points. There is a neighborhood of
$f^{-1}(0)$ of small balls so that $f:\partial(\text{ball})\to{\Bbb R}^n-0
\cong S^{n-1}$.  Now, in each of these small balls, $f$ has either degree
$1$ or $-1$.  If degree equals $1$, then $f|\partial(\text{ball})$
is homotopic to the identity.  If degree $=-1$, then $f|\partial(\text{ball})$
is homotopic to
reflection about the equator.  In these cases $\TI(V^f|\text{ball})=\pm 1=
deg\ f|\partial(\text{ball})$.  Now
$$
\aligned
\TI(V^f) &= \sum\ \TI\ V^f|(\text{ball})\qquad \text{ by proper otopy}\\
&= \sum deg\ f|\partial(\text{balls})= deg\ f'.
\endaligned
$$
\enddemo

\noindent
(12)\quad Suppose $f:M\to {\Bbb R}^n$ where $M\subset{\Bbb R}^n$ is a
codimension zero compact manifold.  Define $V_f(m)=m-f(m)$.  Then
$\TI\ V_f=$ fixed point index of $f$ (assuming no fixed points on $\partial M$)

\demo{Proof}
The fixed point index is defined to be the degree of the map $m\to\frac{m-f(m)}
{\Vert m-f(m)\Vert}$ from $\partial M\to S^{n-1}$.  Hence by (11) we have
the result
\enddemo

\noindent
(13)\quad Let $f:M\to N$ where $M$ and $N$ are Riemannian manifolds and
$f$ is a smooth map.  Let $V$ be a vector field on $M$.  Define the pullback
vector field $f^*(V)$ by
$$
\langle f^*V(m),\vec v_m\rangle =\langle V(f(m)),f_*(\vec v_m)\rangle.
$$
Then if $f:M^m\to {\Bbb R}^n$ so that $f_*|\partial M$ has maximal rank
and $f(\partial M)$ contains no zeros of $V$, then
$$
\TI\ f^*V=\sum v_iw_i+(\chi(M)-deg\ \hat N)
$$
where $v_i=\TI(x_i)$ where $x_i$ is the i${}^{\text{th}}$ zero of $V$, $w_i$
is the winding number of $f|\partial M$ about $x_i$, and $\hat N:\partial M\to
S^{n-1}$ is the normal (or Gauss) map.

\demo{Proof}
In paper \cite{G${}_5$}.
\enddemo
\vfill\eject

\vskip.75truein
Purdue University

\enddocument